\RequirePackage{amsmath}
\documentclass[runningheads, envcountsame, a4paper]{llncs}
\usepackage[T1]{fontenc}
\usepackage[utf8]{inputenc} % colocado por SB
\usepackage[hyphens]{url}            % simple URL typesetting
\usepackage[hidelinks]{hyperref}       % hyperlinks

\usepackage{amsfonts}       % blackboard math symbols
\usepackage{nicefrac}       % compact symbols for 1/2, etc.
\usepackage{microtype}      % microtypography

\usepackage{graphicx}
\usepackage{booktabs}
\usepackage{latexsym}
\usepackage{longtable}
\usepackage{float}
\usepackage[colorinlistoftodos,prependcaption,textsize=tiny]{todonotes}
\usepackage{geometry}
\usepackage{marginnote,letltxmacro}

\LetLtxMacro\mn\marginnote
\usepackage{csquotes}
\usepackage{enumitem} %space lists: use [noitemsep]
\setlist{noitemsep} % to leave space around whole list
\usepackage{algorithm}
\usepackage[noend]{algpseudocode}
\usepackage[gen]{eurosym}
\begin{document}

\title{Generational political dynamics of\\
retirement pensions systems%\thanks{Grants or other notes
%about the article that should go on the front page should be
%placed here. 
%General acknowledgments should be placed at the end of the article.}
}
\subtitle{An agent based model}

\titlerunning{Dynamics of pension systems}       % if too long for running head

\author{Sérgio Bacelar\inst{1} \and Luis Antunes\inst{2}}

\authorrunning{S. Bacelar and L. Antunes} % if too long for running head

\institute{Lisbon University Institute ISCTE-IUL, Lisbon, Portugal 
\email{smbsa@iscte-iul.pt}\\ \and 
GUESS/BioISI — Instituto de Biosistemas e Ciências Integrativas, Faculdade de
Ciências, Universidade de Lisboa, 1749-016 Lisboa, Portugal 
\email{xarax@ciencias.ulisboa.pt}
}
%\date{\today}
% The correct dates will be entered by the editor

\maketitle

\noindent
\makebox[\linewidth]{\small \today}
%%%%%%%%%%%%%%%

\begin{abstract}
  The increasing difficulties in financing the welfare state and in particular
  public retirement pensions have been one of the outcomes both of the decrease
  of fertility and birth rates combined with the increase of life expectancy.
  The dynamics of retirement pensions are usually studied in Economics using
  overlapping generation models. These models are based on simplifying
  assumptions like the use of a representative agent to ease the problem of
  tractability. 
  
  Alternatively, we propose to use agent-based modelling (ABM), relaxing the
  need for those assumptions and enabling the use of interacting and
  heterogeneous agents assigning special importance to the study of
  inter-generational relations. We treat pension dynamics both in economics and
  political perspectives. The model we build, following the ODD protocol, will
  try to understand the dynamics of choice of public versus private retirement
  pensions resulting from the conflicting preferences of different agents but
  also from the cooperation between them. The aggregation of these individual
  preferences is done by voting.  We combine a microsimulation approach
  following the evolution of synthetic populations along time, with the ABM
  approach studying the interactions between the different agent types. 
  
  Our objective is to depict the conditions for the survival of the public
  pensions system emerging from the relation between egoistic and altruistic
  individual and collective behaviours.
\end{abstract}

\section{Introduction}
\label{intro}
Retirement pensions remains an important issue due to the increasing
difficulties to finance the welfare state. The decrease of fertility and birth
rates combined with the increase of life expectancy contribute to population
ageing, which has as a consequence an increase of dependency rates: fewer
workers have to pay pensions for more old retirees~\cite{ec2015}.

These problems have led governments all over the world to design and implement
parametric or structural reforms on pension systems.

To understand these reforms and generally the impact of population ageing on
pension systems, it is necessary to take into account the voter's preferences
and the underlying political environment~\cite{pamp2015}. Voters choose policy
alternatives by election, lobbying, or by campaigning to influence the
Government.

In the case of retirement, different age groups. young, older workers, and
retirees, all have different preferences about pension policies. The direction
of policy reform, results from the aggregate of individual preferences. One of the ways to accomplish this aggregation is through the electoral process.

As such, we argue that pension policy is mainly a political problem and any
reform of a pension scheme, being parametric or non-parametric, is
redistributional~\cite{pamp2015}, involving necessarily winners and losers and
by that reason is a politically very contested issue.

Our aim is to examine the intra and intergenerational political dynamics of
retirement pensions systems, using agent-based modelling (ABM). We will try to
depict the dynamics of choice of public versus private retirement pensions.
These dynamics have been initially studied on the basis of \emph{overlapping
generations} (OLG) models~\cite{samuelson1958}. 

Using ABM to study the the dynamics of retirement pensions enables to understand
the evolution of agents according to age and also to the generation to which
they belong, over time. Usually in economics, OLG models are based on
assumptions that are not always realistic, to ease the problem of tractability.
One of the advantages of ABM is to be able to relax them. Relaxing these
assumptions is expected to ease the understanding of complex processes and to
produce more realistic results. Within ABM it is possible to include in the
analysis multiple variables as the rate of electoral participation (one of the
ways of exercising or influencing power) by age and other variables, or the
existence of income inequalities and savings between younger, older workers and
pensioners. Or even, to contemplate personal bonds between particular
individuals in different generational groups (e.g.\ individuals belonging to the
same family), that could harbour and foster particular positions not similar to
the preferences that would otherwise exist between anonymous agents.

We build a model that accounts for the resilience of retirement pensions
systems, mainly of unfunded pensions, and the persistance of a norm that implies
that \textquote{young workers agree to pay the pension of
retired people in return for the promise that the next generation of workers
will pay for their pension}\cite{blake2006}.

\section{Literature review}
\label{lit}

\subsection{Objectives, reasons and schemes of retirement pensions}

  There are two main objectives of retirement pensions: the first, called
  ``piggybank'' function of pensions, aims to provide insurance against low
  income and wealth in old age. It is a mechanism of consumption smoothing
  across lifetime. The second, the ``Robin Hood'' function, aims to relieve
  poverty and redistribute income and wealth~\cite{barr2008,barr1998,blake2006}.
  
  Different reasons justify the introduction of some form of welfare
  arrangements: the impact of urbanization, industrialization and the demise of the extended family; the extension of voting rights or the increasing strength of left parties and labor unions~\cite[p. 55]{pamp2015} and also long-term change of values related to increased democratisation and the value of all human lives.

  There are two main schemes of pension arrangements: a PAYG (pay-as-you-go)
  system where current workers pay current retirees' pension through a tax that
  is levied on their working incomes: and a pre-funded system, when
  contributions are invested in assets that are accumulated as capital. Benefits
  for retirees result from a previous generated stock of funds.

  Pension contributions can be mandatory or voluntary. The need for a mandatory
  public system is usually justified by two reasons:

\begin{itemize}
    \item the first reason is myopic individual saving behaviour, since
    short-sighted individuals do not save sufficiently for their retirement.
    Individuals tend to revise their consumption plans in an inconsistent way by
    using a higher discount rate for the near future than the far future.
    \item the second reason has to do with imperfect financial markets.
    Private financial markets do not provide sufficient possibilities for
    annuitization of pension benefits due to ``adverse selection
    problems''~\cite{diamond2004} i.e.\ because of asymmetric information.
    \textquote{In most countries private markets for life annuities are
    rudimentary, probably because, where life annuities are not mandatory, the
    individuals who buy them tend on average to be wealthier and to have longer
    life expectancies}\cite{simonovits2003}.
\end{itemize}

  These reasons would justify a paternalistic government intervention.

\subsection{Pensions in the Economics perspective}
\label{obj}
  
  The \emph{overlapping generations model} (OLG) (Diamond–Samuelson
  Overlapping Generations Model with Certain Lifetimes) has been used to represent the individual's life-cycle behaviour. It deals with the
  aggregate behaviour of individuals and generalizes the lifecycle model. The
  original OLG model was developed by Samuelson~\cite{samuelson1958} and
  Diamond~\cite{diamond1965}.
  
  In this model, in each period \(t\) it is considered that two or three
  generations are alive: workers and pensioners on the simplest version of OLG
  and young workers, old workers and pensioners, on the three generations
  version. This simplification of the number of cohorts existing in each period
  makes the analysis tractable. The assumption of considering two or three
  ageing homogeneous groups has consequences: emphasizing age groups and not
  individuals; presupposing group homogeneity behaviour and not heterogeneity.
  These modelling restrictions are not a valid limitation for an ABM approach,
  since it is possible to simulate populations with a distribution of ages
  closer to the reality of any group or country.

  The relationship between age groups on retirement pensions may be seen through
  the lens of political theory.
  
  \begin{quote}
    These kinds of models show that coalition building might be important. The
    literature has suggested that retirees and older workers may conspire, and
    that an elaborated inter-generational punishment mechanism ensures
    sustainability of the system once it has been introduced.~\cite[p.
    15]{pamp2015}
  \end{quote}
  
  Some authors from the field of political economy even support the thesis of a
  ``grey conspiracy'', i.e.\ a conspiracy of pensioners and older workers,
  against the interests of the young~\cite{pamp2015,mendes2011}. Nevertheless
  it is also known that sometimes older people is altruistic to the younger
  generations providing the young that are outside the system due to
  unemployment or low wages.
  
  Inter-generational behaviour can also result from intra-generational
  distribution of attributes: intra-generational inequality may arouse altruism
  from workers towards old people and, for that motive, reinforce public pension
  systems.

  \subsection{Pensions in the Political Science perspective}

  The existence of different demographic and economic scenarios convey different
  individual preferences resulting in different economic policies~\cite{galasso2017}. The aggregation of individual preferences is done through the electoral process using two different types of electoral institutions:
  direct democracy or a representative democracy.
  
  In the first case, voter's preferences are aggregated by direct referendum.
  The decisive variable is the population growth rate as it determines which age
  group contains the \emph{median voter}~\cite{pamp2015}. This scheme implies
  the simplifying reasoning that decision depends solely on self-interest, which
  in turn is determined by age and work status. There are three possible cases:
  
  \begin{enumerate}
  \item [a)] if the median voter belongs to the young age group then, pre-funded
    pension systems based on individual savings would be favoured;
  \item [b)] if it belongs to retirees, a pay-as-you-go (PAYG) system will be
    maintained or reinforced, even if it would be necessary to change the
    contribution rate; and
  \item [c)] finally, the position of old workers about pension reform depends
    on the size of the existing pension scheme (more generous pensions implies decisions similar to those of the retirees).
  \end{enumerate}
  
  In the second case, in which preferences are aggregated within a
  representative democracy, political parties and the electoral system becomes
  important. In the framework of a probabilistic voting model, where voters are
  imperfectly informed about candidates, and candidates are also imperfectly
  informed about the utility preferences of voter's preferences, the number of
  swing voters becomes the decisive variable~\cite{pamp2015,lindbeck1987}. If
  the political representation is proportional, parties compete for swing
  voters. Thus, the age group with the highest number of swing voters would see
  their preferences respected by parties that want to win the election.
    
  Parties normally favour the preferences of the old in their electoral
  programmes, because old people are not only more ``singleminded'' (meaning
  that their core interest is the value of pensions), but they are also
  ideologically more homogeneous~\cite{pamp2015}.

\medskip
  Some studies explain why a voting majority favours a PAYG pension system
\cite{samuelson1958,aaron1966}. The basic idea is that in a dynamically
inefficient economy, a situation characterised by the possibility of making one
generation better off without making any other generation worse off, a PAYG
saving device would improve the overall welfare if the rate of interest \(r\) is
smaller than the population growth rate \(n\). In this situation, the internal
rate of return \(i\) of a PAYG system is higher than the real return of capital
accumulation.

The internal rate of return (IRR) measures an investment’s rate of return. It is
named internal because it excludes external factors, such as inflation. The IRR
is the \textquote{discount rate that equates the present value of pension
benefits with the present value of contributions}\cite{blake2006}.

In the case of a PAYG system, the investment is the negative payment flow of
contributions and the pay-off of this investment is the positive flow of the the
benefits received during the pension phase. As such the IRR is
\textquote{measured as the proportion of the size of benefits to the size of
contributions}\cite{wilke2005,wilke2009}.

The internal rate of return from this scheme (\(i\)) depends on the product of
the product of the working population growth rate \(n\) by the growth rate of
wages \(w\), according to the classical equation~\cite{samuelson1958}:

\begin{equation}
  1+i=(1+n)(1+w)>1+r
\end{equation}
where,

\begin{enumerate}[label={}]
  \item $i$ = Internal Rate of Return
  \item $n$ = Population growth rate
  \item $w$ = Wages growth rate
  \item $r$ = Rate of interest
\end{enumerate}

Under this situation everyone would be better off contributing to the PAYG
system than to invest in assets. For this reason, a system like this is
unanimous and undisputed by all voters.

However, if the assumption of dynamic inefficiency is disregarded, this voter's
behaviour becomes far from being consensual, since in this case it is possible
that some generation(s) would be better off than others.

In the political economics framework of the OLG model one can find usually
several assumptions like the existence of an imperfect capital market, or the
voter's decisions being dealt with like a one-shot vote, binding all future
generations~\cite{pamp2015}. There are even other assumptions, such as the fact
that voters are not altruistic towards other generations, people of the same age
group being homogeneous, with the same preferences and sharing the same utility
function:

\begin{equation}
  U_t = u[c_t^y]+u[c_{t+1}^o]+u[c_{t+2}^r]
\end{equation}
where, for example, $u[c_{t+1}^o]$ is the utility of consumption ($c$) of older workers ($o$) in period $t+1$.

Other assumptions used traditionally in Economics include: the use of a
\emph{representative agent} (by generation) as a typical decision-maker; that
decision on Social Security is done by majority voting; that population growth
has a constant rate \(n\); and also that wage rate is exogenous and constant.

None of these assumptions are mandatory when using a ABM approach, since it
is possible to simulate any level of heterogeneity for each generation.

Another simplifying assumption is that \textquote{utility in
every period of life is only dependent on the level of personal consumption
\(c\)}\cite{pamp2015}.

According to the usual OLG notation, in each period \(t\) there are \(N_t^i\)
individuals, where the superscript \(i \in \{y,o,r\}\) denotes the generation of
young workers, old workers or retirees respectively. One important and
restrictive assumption is that every generation \(N_i\), and the population as a
whole, grow with a constant rate \(n\)~\cite[p. 54]{pamp2015}:

\begin{equation}
  N^y_t=(1+n)N^y_{t-1}
\end{equation}

Consumption during working age and after retirement is determined by wage rate
\(w\), saving rate \(s\) and contribution rate to the pension system \(\tau\).
\emph{Working age} individuals divide their income between consumption, savings
and contributions to be transferred as lump-sum to retirees. \emph{Pensioners}
derive their retirement consumption from the public pension \(x\) and the
accumulated savings, which earn an interest rate \(r\)~\cite[p. 55]{pamp2015}. These are all simplifying assumptions that can be deepened and nuanced in subsequent analysis~\cite{hassan2008}.

\medskip
Some of the limitations of the OLG model are the use of two or three
generations, and not individuals with some age heterogeneity and the use of a
\emph{representative agent} as a typical decision-maker, as well as considering
that people of the same group are homogeneous, i.e.\ have the same preferences,
and treating age rate and population growth as exogenous and constant.

As the only choice variable is the contribution rate to pensions ($\tau$), the
result of the vote on pension system \textquote{depends on
the preferences of the three generations and their relative sizes}\cite{pamp2015}.

In this model young workers only support the introduction of a PAYG scheme if
the population growth rate exceeds the interest rate in a economy dynamically
inefficient. Retirees only have the contribution rate as a choice variable. For
older workers, we can say that the closer an individual is to retirement, the
more profitable the introduction of a public pension system becomes. We can
conclude that \textquote{both old workers and
retirees vote in favor of a PAYG scheme}\cite[p. 57]{pamp2015}.

The preference ordering of the three generations is therefore:

\begin{equation}
  \tau^r = 1 > \tau^o > \tau^y
\end{equation}

Assuming a direct majority vote, the relative sizes of the three age groups and
the identity of the median voter will dictate the outcome. Contradicting this model that presupposes an equivalence between group and voter size, it is possible using the agent-based model alternative perspective to take into account that vote turnout rate diverge by age and so the identity of the median voter may be relatively independent of the age group size. As voter turnout is normally positively related to age it can reinforce the probability of older median voter.

In the case of pension politics as a repeated voting game, we can ask ourselves
what makes self-interested individuals introduce and maintain a pension system.
It could be the result of a social contract between generations, enforced by
either through punishment or reputation.

It seems that other factors, such as majority voting combined with altruism or
intra-generational heterogeneity, would give better explanations to the
political sustainability of public pension systems. But altruism does not generalize to culturally different countries.

\medskip
PAYG pension systems not only transfer resources from the young to the retired,
but also from the better off to the worse off, contributing to some
intra-generational redistribution. This is one of the main keys to understand
why the preference for public pension systems is income inequality, measured for
example by the worker's wage relative to the average income.

Pension systems, are commonly divided into two broad classes: they are organized
according to the principles of either the Beveridgean or the Bismarckian
tradition. Conceptionally, a Bismarckian pension system is characterized by a
close link between previous earnings (and contributions when we assume that the
latter are collected as payroll taxes) and today’s benefits. A Beveridgean
pension system, on the other hand, provides a basic or minimum pension.

The resulting measure of the level of intragenerational redistribution in the
public pension system may be referred to as the Bismarckian factor. 

In a ``pure'' Beveridgean pension system, every pensioner receives the same
pension benefit, independent of his/her (previous) household income. Here, the
Bismarckian factor assumes a value of zero. 

Under a ``pure'' Bismarckian pension system, benefits are proportional to
previous earnings/con\-tributions, i.e.\ pension benefits exhibit the same level
of inequality as earnings. Accordingly, the Bismarckian factor equals one.

Let \(Y^i\) and \(P^i\), denote the mean income and the mean pension benefit,
respectively, of the $i^{th}$ quintile of the income distribution ($i \in
{B(ottom),2,3,4,T(op)}$). A purely Bismarckian pension system implies
$\frac{P^B}{Y^B} = \frac{P^T}{Y^T}$, and a purely Beveridgean pension system
implies $P^B = P^T$. The pension benefit of a representative member of quintile
\(i\), \(P^i\), is defined as a convex combination of a flat payment
(proportional to the mean income) and an earnings-related component
(proportional to \(Y^i\)):

\begin{equation}
  P^i \equiv \tau [\xi Y^{i}+(1-\xi)\mu]
\end{equation}
where \(\xi \in [0, 1]\) is the Bismarckian or redistribution factor \(\mu = \sum_i Y^{i}/5\) is mean income of a society and \(\tau \equiv \sum_i P^{i} / \sum_i Y^{i} \in [0,1]\) the `generosity index', a measure of the level of redistribution between generations~\cite{krieger2013}.

The higher \(\xi\) the less intra-generational redistribution takes place.
(\(\xi = 0\) indicates a perfect Beveridgean system).

\section{Model}
To define the model we use, we partially follow the ODD (Overview, Design
concepts, Details) protocol~\cite{grimm2006}.

\subsection{Purpose}
This model will try to show the dynamics of choice of public versus private
retirement pensions. It will try to explain the resilience of retirement
pensions systems, mainly of unfunded pensions.

Our objective is to to study variation of favorability of heterogeneous agents
to public pensions (PAYG).

Some of the hypotheses to test are:
\begin{enumerate}
\item preferences for a big public PAYG system increase with age;
\item older workers policy preferences depend on population growth and on the
size of the existing pension scheme.
\end{enumerate}

In order to evaluate the favorability to public pensions it is possible to compute cohort--specific rates of return. Individuals that belong to cohorts that due to the fact of evaluating positively the sum of their contribution payments are expecting greater internal rates of return may have a greater chance to be more favorable to public pensions. Rates of return can be computed for different demographic groups (single men, single women and married couples) and different scenarios (retirement at the statutory retirement age, retirement at earlier ages etc.). This is a deterministic approach, contrasting with a stochastic approach where the expected payment flows includes longevity, survival risks as well as the time of retirement.

To calculate the rates of return we have to consider the flux of (negative) contribution payments and (positive) pension benefits to be discounted to a common date, which is the date of entry to the labor force. The internal rate of return is the rate at which the net present value of benefits received, equalizes the net present value of contributions paid:

\begin{equation}
\sum_{a=a_{RA}}^{a_{max_c}}P_{c,a}\left(\frac{1}{1+r}\right)^{a-a_0} = \sum_{a=a_0}^{a_{RA}-1}C_{c,a}\left(\frac{1}{1+r}\right)^{a-a_0}
\end{equation}
Where:
\begin{align*}
  a         &= \text{Age index} \\
  a_0       &= \text{Age of entrance into the labour force} \\
  a_{RA}    &= \text{Retirement age} \\
  a_{max_c} &= \text{Maximum age/end of pension period of cohort \(c\)} \\
  P_{c,a}   &= \text{Pension payments to cohort \(c\) at age \(a\)} \\
  r         &= \text{internal rate of return} \\
  C_{c,a}   &= \text{Contribution payments by cohort \(c\) at age \(a\)}
\end{align*}

\medskip

It is impossible for agents to have complete knowledge of the internal rate of return, population and wages growth or the rate of interest. In the absence of a situation of absolute rationality, favorability to public pensions and decisions about retirement can only be understood in a situation of limited knowledge and bounded rationality.

To model their beliefs agents are classified into three categories: one minority that adopts quasi-optimal behavior, another minority that adopts random behavior, and a majority of agents who mimic members of their social networks~\cite{axtell1999}, social circles~\cite{hamillgilbert2016,hamill2009} or contexts~\cite{nunes2015}. 

\subsection{Entities, state variables, and scales}

\subsubsection{Agents}

The agents in the model will be composed by: individuals (\(i\)); Government (\(G\)); and two competing political parties (\(P_L\) and \(P_R\)).

\paragraph{Individuals}

These are the model parameters of individuals:
\begin{itemize}[noitemsep]
\item age: discrete value \(\in [0;100]\)
\item age-category: four ordered generations (children, young (workers);
adult; old);
\item Gender (male, female), because there are several variable gender related,
as employment rate, civil status, life expectation and couple coordination in
time to retirement.
\item Employment status: inactive, employed/unemployed and pensioner;
Note: all individuals aged 65+ are pensioners.
\item altruism towards the old (\(\beta_i\ \in [0;1]\))
\item pension (\(p_i\)= Reference earnings $\times$ Pension Accrual Rate
$\times$ Sustainability Factor) (monthly)
\item individual \(i\) ideological bias (\(\mu^i \in [0;1]\)): propensity to
vote for party x.
\item political influence of workers and pensioners (\(\phi \in [0;1]\))
\item contribution rate to the pension system (Social Security)(\(\tau = 0.11w \) (gross wage rate)). For self-employed individuals 29,6\%.
\item consumption rate (\(\bar{c}\))
\item saving rate (\(\bar{s}=5.6\%\))
\item wage rate (\(\bar{w} = 17240~\euro{}\) by year) or Income (\(I\))
(distribution of income depends on age) [make income and wage equivalent in a
first phase];
\item welfare favorability (towards retirement public pensions)
(\(0\leq{wf}\leq{1}\)) i.e.\ probability of voting for policies/parties
favorable to welfare state (older workers are ideologically more homogeneous
than others~\cite{pamp2015});
\item choice of type of pension regime: PAYG or private/funded. 
\end{itemize}

\paragraph{Government}
is another entity which can change contribution rates to the public pension
system (\(\tau\)) according to individual income (\(I\)); it also pays (public)
pensions: this implies the need for a computation mechanism of a greater or
lesser favorability to use pensions as redistribution. It can also change wages
in public \(w_i^{pu}\) and private \(w_i^{pr}\) sectors [a possibility is to
treat private wages as a function of public wages].

\paragraph{Political parties}

To simulate aggregation of preferences that use indirect voting (political
parties), we simplify using only two political parties (\(P_L\) and \(P_R\)).
One of the attributes of both parties is common popularity (\(\delta\)), which
is operationalized as a random variable:

\blockquote{It is conceptualized as a random variable, which is uniformly
distributed on (\(\frac{-1}{2d_\delta}, \frac{1}{2d_\delta}\)), has a mean of
zero and cannot be controlled by the competing parties. It is being realized
after all policy platforms have been chosen and therefore captures all the
imponderables of electoral competitions where scandals may be revealed or
foreign policy events (war, terrorist attack)
intervene.~\cite[p.92--93]{pamp2015}}

\subsubsection{Global aggregate variables}

Auxiliary variables were added for monitoring characteristics of the whole
population:
\begin{itemize}[noitemsep]
\item dependency-ratio (\(dr=(young+old)/adults\)) depends on the level of the components. The value for Portugal in 2015 was 53.5\%. Young = people < 15 years; Old = 65+ years;
\item population (natural) growth rate (\(n = -0.22\%\) (2015))
\item wage growth rate (\(W\))
\item real interest rate (\(R\))
\item relative political influence of workers (\(\phi \in [0;1]\))
\item transfers between adults and young/old (PAYG scheme): \(T=I_r\), where
\(r=rate\)
\item redistribution factor relating contributions to benefits (\(\xi \in [0,
1]\), where \(\xi=0\) is a perfect Beveridgean (high redistribution) and
\(\xi=1\) is a Bismarckian situation (low redistribution)
\end{itemize}

\subsubsection{Spatial units}

A two-dimensional \({L}\times{L}\) square lattice; $L=100$ for $n=1000$ agents.

\subsection{Process overview and scheduling}

Agent-based modelling is used to simulate individuals that make decisions that
can impact on the retirement system as lobbying and voting (direct voting or
using parties) for different systems~\cite{klabunde2015a,lovelace2016}.

\subsubsection{Demography}

The demographic characteristics of population are fundamental to define the contours of the mechanisms of the simulation.

A demographic microsimulation is run parallel to the agent-based simulation. The microsimulation determines for each individual all demographic states that an individual will experience throughout his life and the amount of time the individual will spent in each state. See Table~\ref{table:state_space} for state variables and respective possible values.

\begin{table}[htbp]
  \centering
  \caption{State space: State variables and their corresponding values}
  \label{table:state_space}
  \begin{tabular}{@{}ll@{}}
  \toprule
  \textbf{State Variable} & \textbf{Possible Values} \\ \midrule
  Sex & Female (f); male (m) \\
  Marital status & never married (NM); married (M); divorced (D); widowed (W) \\
  Fertility status & childless (0); one child (1); two children (2); three and more children (3+) \\
  Educational attainment & no education (no); primary education only (low); lower secondary school (med); \\ & upper secondary or tertiary education (high) \\
  Employment status & inactive (education) (ie); employed (e); unemployed (u); retired (r) \\
  Mortality & Alive; dead (dead) \\ \bottomrule
  \end{tabular}
\end{table}

It is possible to define a matrix of transition states (and probabilities of transition between states). Examples of transitions:

\begin{itemize}
  \item \emph{Marital status}: NM \(\rightarrow\) M; M \(\rightarrow\) D; M \(\rightarrow\) W; D \(\rightarrow\) M; W \(\rightarrow\) M (marriage1, marriage2, divorce, and widowhood rates)
  \item \emph{Fertility status}: 0\(\rightarrow\)1, 1\(\rightarrow\)2, 2\(\rightarrow\)3+ and 3+\(\rightarrow\)3+ (fertility rates)
  \item \emph{Employment status}: ie \(\rightarrow\) e; e \(\rightarrow\) u; e \(\rightarrow\) r; u \(\rightarrow\) r; u \(\rightarrow\) e (employment and unemployment rates)
  \item \emph{Mortality}: alive \(\rightarrow\) dead (absortion state) (death rates)
\end{itemize}

\begin{itemize}
\item Birth: each female in any simulation year can give birth to a child,
with a probability which is year- and age-specific and is reported in a file.
\item Employment: all individuals who are of working age (>15) and whose
previous work state was neither student nor retired are considered to be
available to work. Conditional on this, individuals are employed with a
probability which depends on age, lagged work state (either employed, unemployed
or inactive): regression coefficients are stored in a file.
\item Death: death is also a probabilistic event, with year and age-specific
death probabilities contained in a file. All individuals die at age 100.
\end{itemize}

One time step represents one year, and simulations will be run for 50 years.

\subsection{Initialization}
Initial populations from two countries, starting with Portugal (using sample
databases from IPUMS -- Minnesota Population Center International Database
Project -- or the OECD database:~\cite{oecd2015}) with different welfare state
systems (pension systems) will be used. The features of the countries are based
on real countries. The initial population is a synthetic population normally
used on microsimulation studies (see R package MicSim~\cite{zinn2014}). 

Some population features that it will be necessary to monitor during the study:
\begin{itemize}[noitemsep]
  \item Distribution of the resident population by age (years) in 2015;
  \item Distribution of birth rate by age of mother/father
  \item Distribution of mortality rate by age and gender
  \item Migration (will not be taken into account)
  \item Distribution of active population by employment status and age
  \item Distribution of income by age
  \item Education level
  \item Population growth rate (contrary to what happens in the OLG models, does
not change at a constant rate), aging rate, interest rate
  \end{itemize}

\subsection{Action and mechanisms}
\begin{itemize}
  \item Initialization
  \begin{itemize}
    \item[--] Generate a population of 1000 individuals with demographic characteristics similar to country \(X\) on time \(t\)
    \item[--] Generate the Government entity and two Parties entities
  \end{itemize}
  \item Demographic schedule
  \begin{itemize}
    \item[--] Each year the values of state variables of each individual are affected by the transition rates
  \end{itemize}
  \item Redistribution
  \begin{itemize}
    \item[--] Each year individuals transfer a percentage of annual income
    (contributions to Social Security) to Government or/and save it to private
    funding
    \item[--] Each year Government distributes the sum of collected contributions to
    pensioners
  \end{itemize}
  \item Voting
  \begin{itemize}
    \item[--] Each four years individuals vote
  \end{itemize}
\end{itemize}

The policy decisions (pension parametric reforms) of the two parties are the following:

\begin{itemize}
  \item $P_L$ increases contributions to Social Security ($\tau$) and/or decreases retirement age and/or increases pensions;
  \item $P_R$ decreases contributions to Social Security ($\tau$) and/or increase retirement age and/or decreases pensions.
\end{itemize}

\section{Concluding Remarks and Further Research Directions}

The model presented above needs to be operationalized, and a reflection of their
results must be done. Nevertheless, it is possible even beforehand to present
some remarks that we think will be useful for further research. 

When a population subset (intermediated or not by a party), wins an election, it can never win on every issue. It will always lose in some aspect. The winner party, favouring public pensions, may simultaneously harm other equally desirable policies (e.g. public transportation, health, children). Public investment is scarce, and choices must be made.

Choosing a party by vote may induce some feedback on every agent because they
all will be affected by choice, one way or another. It is possible to think in
an initially randomly distributed ideology favouring one party or another. This
factor could be changed along with time and circumstances of the simulation.
Even altruism may be considered to be initially randomly distributed in the
simulation. Altruism represents consideration to other people, knowledge and
intergenerational relationship. It works as a point of resistance to selfish
choices.

The distribution of agents by the three generations considered by the OLG model is done as a function of age and work status and not generation belonging. This fact means that it will be regarded as only two generation types: workers (with different ages) and pensioners. Old workers with 65 or more years will be considered as pensioners. 

The level of favorability to Welfare State or the probability to vote for $P_L$ is a stochastic function of age. Younger people have less likelihood of favouring Welfare State. Also, younger people have a less voter turnout level.

In the initialization of the simulation, we can consider that some agents may
know almost correctly the internal rate of return. However, they should be a
minority and distributed randomly.

One of the most decisive factors in the simulation will be income and inequality
level because it has a significant impact on consumption and savings. When there
are not enough savings, there is no motivation for investment on pension funds.
However, income and inequality are not generated by the simulation, being
exogenous parameters, assuming different values in different simulation rounds.

Also, the wage rate ($w$) and interest rate ($r$) will be treated as exogenous, functioning only as parameters in the behaviour space.

It is possible in the model to simulate a more Bismarckian or Beveridgian system
pension, controlling the $\xi$ parameter in the following equation, or the
inter-generational redistribution, controlling the parameter $\tau$.

\begin{equation}
  P^i \equiv \tau [\xi Y^{i}+(1-\xi)\mu]
\end{equation}

Favouring public pensions is a function of the expectation of the internal rate of return. This expectation is an idea, real or unreal that the contributions to pensions by some individual will be fairly rewarded in the future, after retirement. The construction of this idea considers the level of the sum of contributions during all the period until retirement. This construction implies that the age of entry into the labour force, the level of wages and the official contribution rate and the age of retirement, are factors to take into account.

More imponderable factors, like life expectation at the time of retirement and substitution rate, are also decisive factors of the internal rate of return function.

%\begin{acknowledgements}
%If you'd like to thank anyone, place your comments here
%and remove the percent signs.
%\end{acknowledgements}

% BibTeX users please use one of

\bibliographystyle{splncs04}

\end{document}